\documentclass[]{spie}  

\usepackage{amsmath,amsfonts,amssymb}
\usepackage{graphicx}
\usepackage[colorlinks=true, allcolors=blue]{hyperref}

\title{Toward volume manufacturing of high-performance soft x-ray critical-angle transmission gratings}

\author[a]{Ralf K. Heilmann}
\author[b]{Alexander R. Bruccoleri}
\author[a]{Jungki Song}
\author[c]{Matthew T. Cook}
\author[c]{James A. Gregory}
\author[c]{Renee D. Lambert}
\author[c]{Dimitri A. Shapiro}
\author[c]{Douglas J. Young}
\author[d]{Miranda Bradshaw}
\author[d]{Vadim Burwitz}
\author[d]{Gisela D. Hartner}
\author[d]{Andreas Langmeier}
\author[e]{Randall K. Smith}
\author[a]{Mark L. Schattenburg}
\affil[a]{Space Nanotechnology Laboratory, MIT Kavli Institute for Astrophysics and Space Research, Massachusetts Institute of Technology, Cambridge, MA 02139, USA}
\affil[b]{Izentis, LLC, Cambridge, MA 02139, USA}
\affil[c]{MIT Lincoln Laboratory, Lexington, MA 02421, USA}
\affil[d]{Max-Planck-Institut f\"ur Extraterrestrische Physik, 85748 Garching, Germany}
\affil[e]{Center for Astrophysics, Harvard-Smithsonian Astrophysical Observatory, Cambridge, MA 02138, USA}

\authorinfo{Further author information: (Send correspondence to R.K.H.)\\E-mail: ralf at space.mit.edu, , URL: http://snl.mit.edu/}

\begin{document} 
\maketitle

\begin{abstract}
High-resolution ($R = \lambda /\Delta \lambda > 2000$) x-ray absorption and emission line spectroscopy in the soft x-ray band is a crucial diagnostic for the exploration of the properties of ubiquitous warm and hot plasmas and their dynamics in the cosmic web, galaxy clusters, galaxy halos, intragalactic space, and star atmospheres.  Soft x-ray grating spectroscopy with $R > 10{,}000$ has been demonstrated with critical-angle transmission (CAT) gratings.  CAT gratings combine the relaxed alignment and temperature tolerances and low mass of transmission gratings with high diffraction efficiency blazed in high orders.  They are an enabling technology for the proposed Arcus grating explorer and were selected for the Lynx design reference mission grating spectrometer instrument.  Both Arcus and Lynx require the manufacture of hundreds to perhaps $\approx 2000$ large-area CAT gratings.  We are developing new patterning and fabrication process sequences that are conducive to large-format volume processing on state-of-the-art 200 mm wafer tools.  Recent x-ray tests on 200 nm-period gratings patterned using e-beam-written masks and 4x projection lithography in conjunction with silicon pore focusing optics demonstrated $R \approx 10^4$ at 1.49 keV.  Extending the grating depth from 4 $\mu$m to 6 $\mu$m is predicted to lead to significant improvements in diffraction efficiency and is part of our current efforts using a combination of deep reactive-ion etching and wet etching in KOH solution.  We describe our recent progress in grating fabrication and report our latest diffraction efficiency and modeling results.
 
\end{abstract}


\section{INTRODUCTION}
\label{sec:intro}  

 The soft x-ray band covers the characteristic lines of the most abundant ``metals" (O, C, Ne, Fe, N, Si, Mg, S).  Their ionization states and line ratios provide crucial diagnostics about the plasmas they are part of.  High-resolution absorption and emission spectra tell us about hot gas dynamics at the edges and farther outside of galaxies, revealing how galaxies have formed and evolved, and how supermassive black holes at the center of galaxies interact with their surroundings.  They can help us map the distribution of the missing baryons, presumed to be hiding in the warm-hot intergalactic medium.  On a smaller scale accretion of hot gas during star formation also generates soft x-ray spectra with rich plasma diagnostics. More detailed descriptions can be found in work by Smith\cite{Arcus2019} and Bautz.\cite{Lynx2019}  

 Currently operating x-ray spectroscopy missions
 (the Chandra High Energy Transmission Grating Spectrometer (HETG)\cite{cxc} and the XMM-Newton Reflection Grating Spectrometer (RGS)\cite{RGS}, both launched 20 years ago with technology developed 30 years ago) lack both the effective area and the resolving power ($R = \lambda /\Delta \lambda$) for these measurements.  Luckily, a generation of technology development later, we can now build an instrument on an Explorer budget that is much more powerful than what used to require a flagship mission, as the Arcus\cite{Arcus2019} soft x-ray grating Explorer mission concept shows.  Arcus features a modular design with four parallel optical channels, each consisting of a co-aligned array of 12 m-focal length silicon pore optic (SPO) mirror modules,\cite{Collon2018} developed in Europe for the Athena\cite{Athena} mission.  Just aft of each mirror array follows an array of a few hundred critical-angle transmission (CAT) gratings arranged on the surface of a tilted Rowland torus.\cite{SPIE18,moritz2017,moritz2018}  Arcus will have an effective area of at least 200 cm$^2$ and $R > 2500$ over the 1-3 nm wavelength range, exceeding HETG and RGS figures of merit by factors of at least 5-10.   The soon (2022) to be launched Resolve microcalorimeter on XRISM\cite{XRISM} will also be far from competitive with Arcus performance in high-resolution soft x-ray spectroscopy of point sources.
 
 The current Astrophysics Decadal Review is evaluating four Surveyor-class mission concepts.  One of them, the Lynx\cite{Gaskin2019} x-ray observatory, features a retractable grating array for large-area ($> 4000$ cm$^2$) high-resolution ($R > 5000$) soft x-ray spectroscopy, covering a fraction of a large x-ray optic. A CAT grating spectrometer made of $\sim 2000$ large-area gratings\cite{CAT-JATIS,moritz2019} was selected for the Design Reference Mission described in the Lynx X-ray Observatory Report.\cite{Lynxreport}  As part of this effort we developed a detailed Technology Development Roadmap that can take CAT grating technology from today's state-of-the-art to Technology Readiness Level (TRL) 6 for Lynx by 2024.\cite{CATroadmap}
 
Technology readiness is a very useful concept regarding the maturation of new technology.  We have previously described CAT grating TRL in more detail.\cite{SPIE19} However, when more than a few manifestations of the technology are needed, the manufacturability of a large number of very similar or almost identically performing items needs to be considered as well.  As with the mass fabrication of tens of thousands of mirror segments for a large x-ray mirror,\cite{metashells} a manufacturing approach must be developed that can reliably produce many hundreds to a few thousands of interchangeable grating elements.  

Below we first give a brief overview of the CAT grating principle, our structural design, and our fabrication approach for individual gratings for technology development.  We describe the recently demonstrated addition of aligned front and back side structures.  This is followed by discussion of process transfer to tools capable of repeatable batch processing of 200 mm wafers, which will greatly automate and accelerate grating fabrication.  We then present recent measurements of x-ray resolving power of a grating patterned with the new tool set and discuss the modeling of diffraction efficiency data before we conclude and summarize.

\section{CAT GRATING PRINCIPLE AND STRUCTURAL HIERARCHY}

\begin{figure} [ht]
   \begin{center}
   \begin{tabular}{ c c } 
   \includegraphics[height=4.5cm]{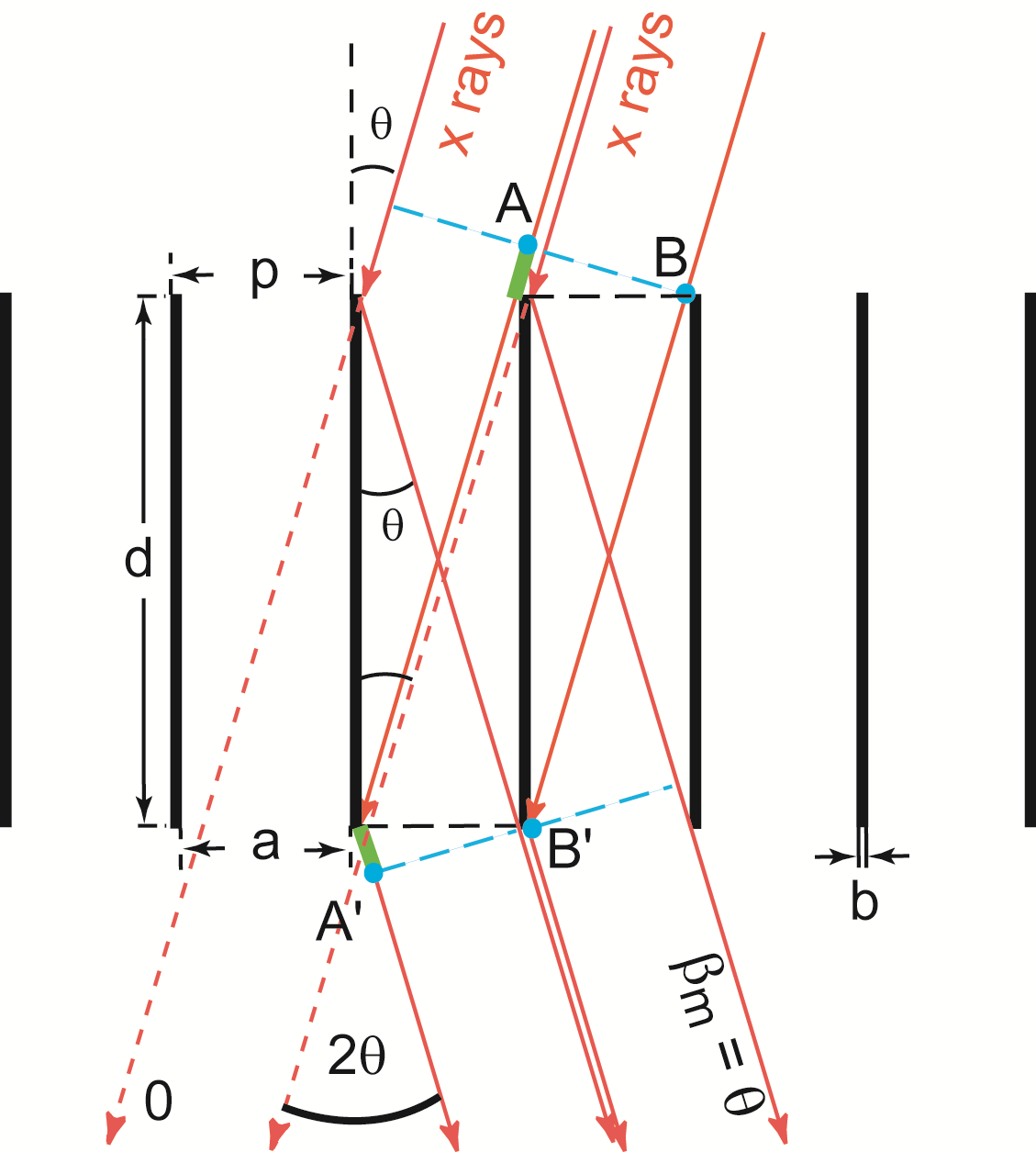}
   \includegraphics[height=2cm]{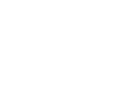}
   \includegraphics[height=4cm]{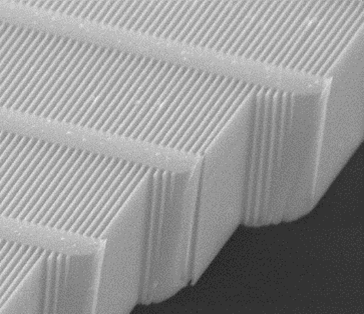}
   \end{tabular}
   \end{center}
   \caption[example] 
   { \label{fig:cross} 
Left: Schematic cross-section through a CAT grating of period $p$. The $m^{\rm {th}}$ diffraction order occurs at an angle $\beta_m$ where the path length difference between AA' and BB' is $m\lambda$. Shown is the case where $\beta_m$ coincides with the direction of specular reflection from the grating bar sidewalls ($| \beta_m | = |\theta|$), i.e., blazing in the $m^{\rm {th}}$ order. Right: Scanning electron micrograph of a cleaved CAT grating membrane showing top, cross-section and sidewall views of the 200 nm-period silicon grating bars and their monolithically integrated 5 $\mu$m-period cross supports (x rays enter from the top and leave out the bottom).   }
\end{figure}

CAT gratings are comprised of ultra-high aspect-ratio, freestanding grating bars with nm-smooth sidewalls.  The gratings are rotated by an angle such that x rays of wavelength $\lambda$ impinge on the sidewalls at graze angles $\theta$ below the critical angle for total external reflection $\theta_c$ (see Fig.~\ref{fig:cross}).  The grating equation provides the $m^{\mathrm {th}}$ order diffraction angle $\beta_m$ via

\begin{equation}
{m \lambda \over p} = \sin \theta - \sin \beta_m ,
\label{ge}
\end{equation}

\noindent
with $p$ being the grating period.  Diffraction orders near the direction of specular reflection from the sidewalls show increased efficiency (i.e., blazing).  The small critical angles for soft x rays (typically on the order of 1-2 degrees) demand high aspect ratio grating bars in order to intercept all incoming photons.  Furthermore, the bars should be as thin as possible to minimize absorption.  The grating period cannot be too large compared to the x-ray wavelength to obtain diffraction orders that can be sorted by order using the energy resolution of Si-based detectors.  We initially chose a design with grating period $p = 200$ nm, grating bar depth $d = 4$ micrometers, and bar thickness $b \approx 60$ nm.  Such a grating is most efficient when $\tan \theta \approx (p-b)/d$, or $\theta \approx 2$ deg.

\begin{figure} [ht]
   \begin{center}
   \begin{tabular}{c  c  c } 
   \includegraphics[height=4cm]{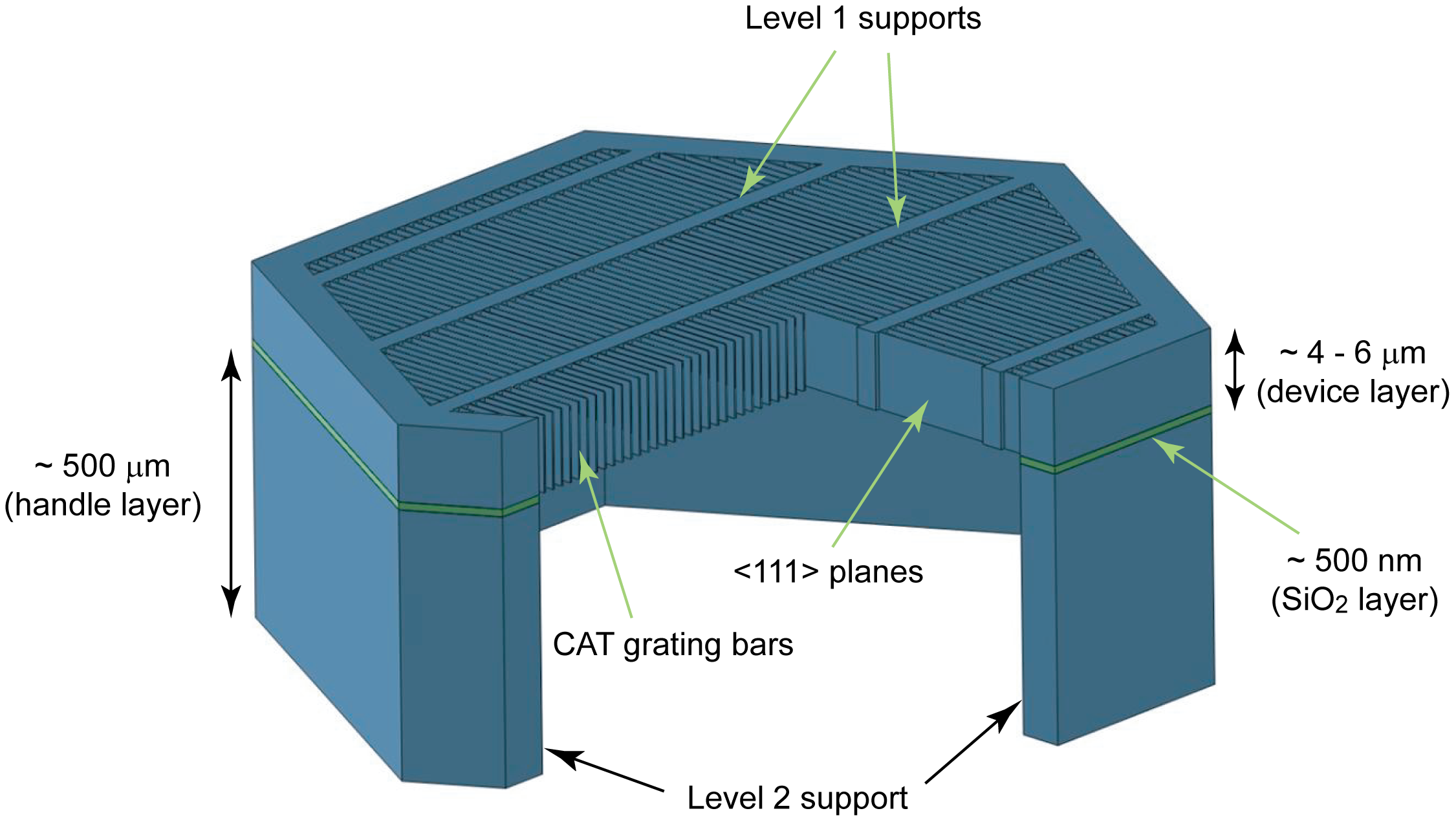}
   \includegraphics[height=4cm]{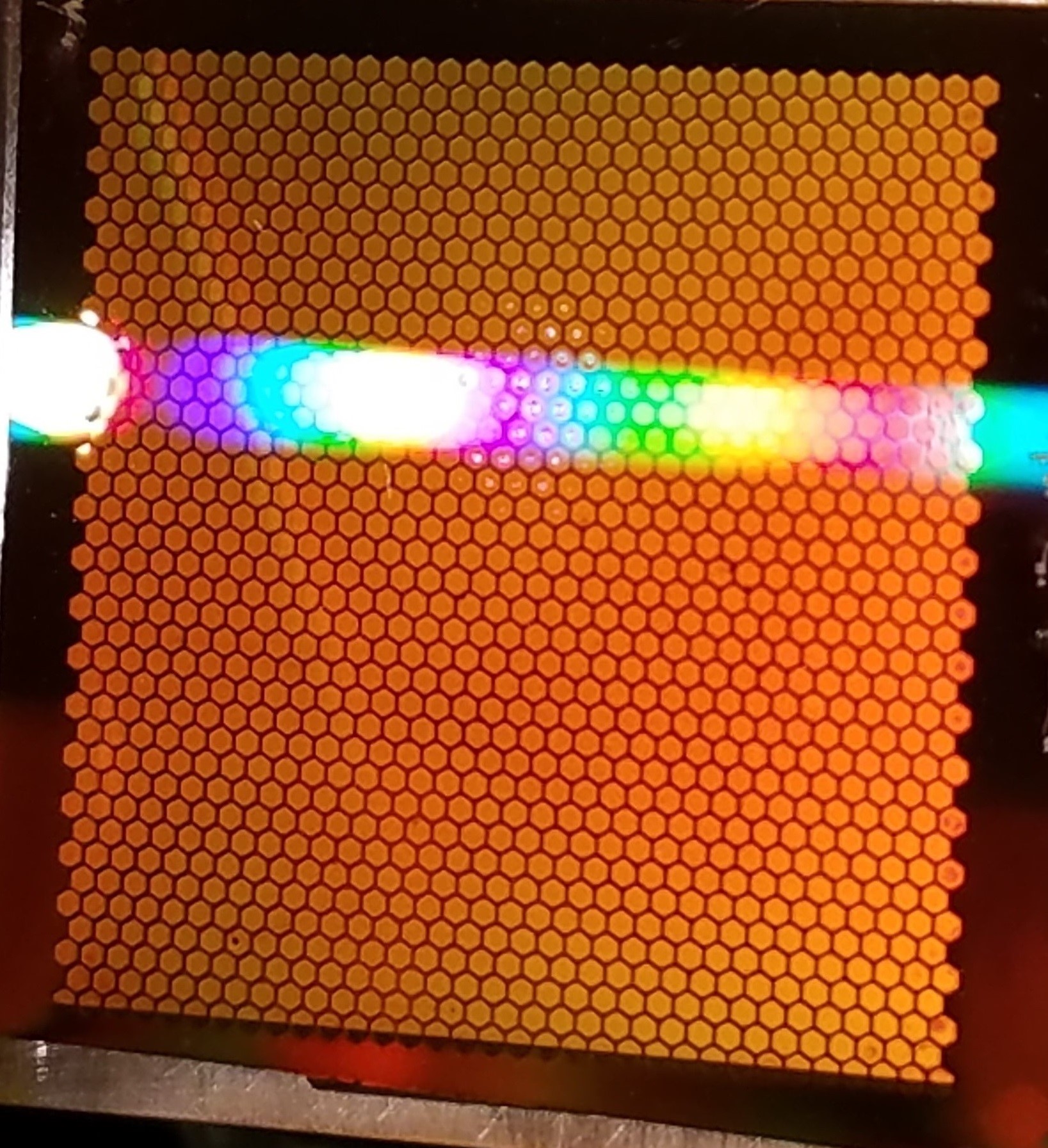}
   \end{tabular}
   \end{center}
   \caption[example] 
   { \label{fig:L1} 
Left: Schematic showing the structural hierarchy of a CAT grating membrane (not to scale). Right: Photograph of an existing 32 x 32 mm$^2$ CAT grating membrane with back illumination to show the hexagonal L2 mesh, and with visible light diffraction due to the L1 mesh.}
\end{figure}

CAT grating bars are not supported by a membrane, but freestanding.  As seen on the right in Fig.~\ref{fig:cross}, the bars are held in place by a monolithically integrated 5 $\mu$m-period Level 1 (L1) support mesh.  In order to manufacture CAT gratings that can cover large areas on the order of thousands of square centimeters, additional support structures are needed for the few-$\mu$m thin grating layer.  Fig.~\ref{fig:L1} shows an additional, much thicker and stronger Level 2 (L2) hexagonal support structure on the scale of $\sim 1$ mm.  The photograph on the right shows a so-called grating membrane, featuring an additional Level 3 (L3) frame around the edge.  This edge is used to bond the membrane to a metal frame that serves as the mechanical interface to a larger machined grating array structure (GAS).

\section{GRATING FABRICATION APPROACH}

\begin{figure} [ht]
   \begin{center}
   \begin{tabular}{c} 
   \includegraphics[height=8cm]{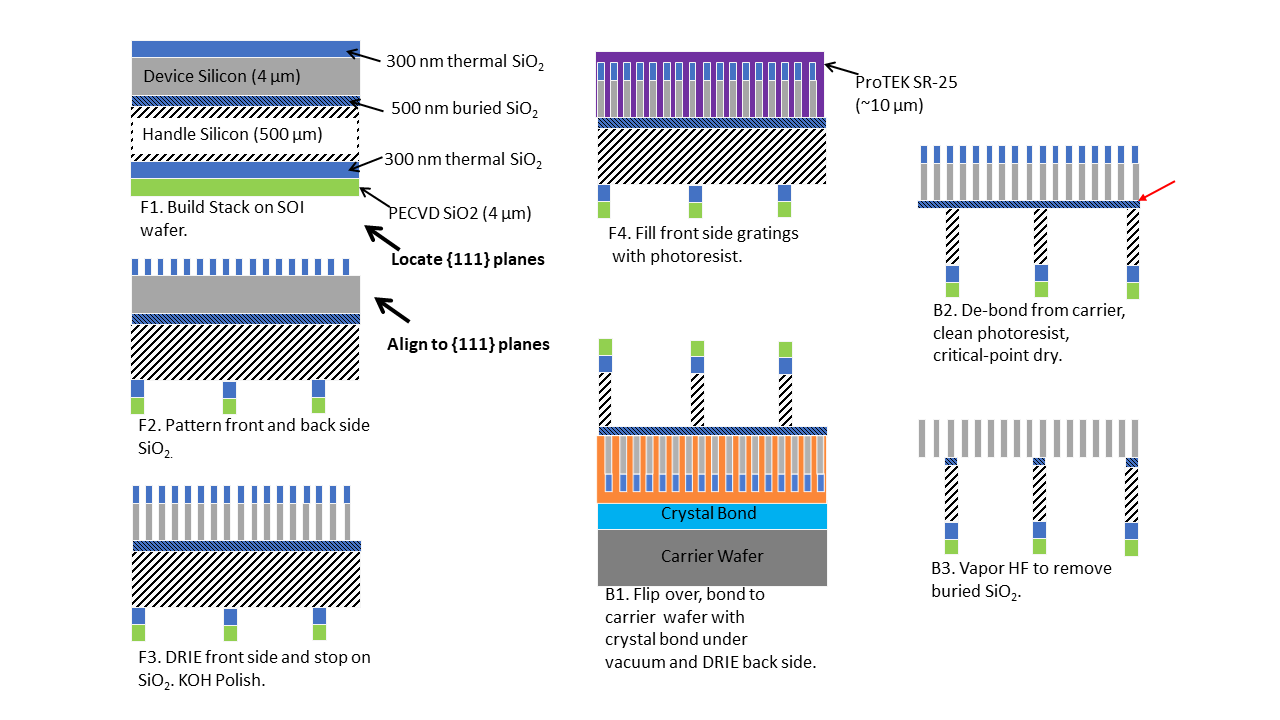}
   \end{tabular}
   \end{center}
   \caption[example] 
   { \label{fig:Fab} 
Highly simplified schematic overview of CAT grating fabrication steps.  See text for more details.}
\end{figure}

We have fabricated numerous CAT gratings in sizes up to $32 \times 32$ mm$^2$ in our lab and other MIT campus facilities, using 100 mm-diameter silicon-on-insulator (SOI) wafers with a (110) device layer (front side) of thickness $d$, a thermal oxide layer on both sides, and an additional plasma-enhanced chemical vapor deposition (PECVD) oxide layer on the handle layer (back side) (see Fig.~\ref{fig:Fab}).\cite{alex2,EIPBN2016}   The grating pattern is defined with interference lithography.  The grating lines are aligned parallel to one set of vertical \{111\} planes. The L1 cross-support mesh is defined in similar fashion.  The combined pattern is transferred into the thermal oxide, which serves as a mask for the subsequent deep reactive-ion etch (DRIE).  The buried oxide (BOX) layer serves as an etch stop.  A short anisotropic wet etch in KOH solution greatly reduces the scalloping from the DRIE and leaves much smoother \{111\} grating bar sidewalls.  The front side is protected and bonded to a carrier wafer.  The L2 hexagon mesh is patterned into the back side PECVD oxide, surrounded by a narrow frame area.  This pattern is etched through the $\approx $ 500 $ \mu$m-thick handle layer using DRIE, again using the BOX layer as an etch stop.  Then the wafer is detached from the carrier, cleaned, and critical-point dried.  The BOX layer is removed from the open areas using RIE and HF vapor etching, resulting in a freestanding grating with integrated L1 cross-supports and a strong L2 mesh that keeps large-area gratings supported (see Fig.~\ref{fig:L1}).

\section{aligned front side L2 mesh}

At the end of the above fabrication sequence device layer and handle layer are only connected by buried oxide in the areas where the front side L1 mesh and the back side L2 mesh and L3 frames overlap (see red arrow in step B2 of Fig.~\ref{fig:Fab}).  During the isotropic HF vapor etch, undercutting can lead to thinning of the connected areas or even detachment between device and handle layers.  It is difficult to time the HF etch perfectly right, and variations in grating bar widths and device layer and BOX thickness may lead to leftover oxide between grating bars in some areas and detachment in others.  A better approach for higher vapor HF process latitude and improved mechanical behavior is to incorporate an L2 mesh in the front side pattern that has the same periodicity and is aligned with the back side L2 mesh.  This requires a third front side lithography sequence.

In order to incorporate a L2 mesh on the front side, the back side pattern included a set of alignment marks near the edge of the wafer. An additional layer of photoresist was applied to the front side after the CAT and L1 lines were patterned into the front-side thermal oxide layer. This photoresist was patterned with a L2 and L3 mesh and aligned to the back side via the alignment marks. The front side was then etched via the standard DRIE and subsequent processing steps depicted in Fig.~\ref{fig:Fab}, starting on step F3. The resulting grating was significantly easier to process and the vapor HF step did not damage the membrane via the undercutting observed without the front-side L2 mesh. Future gratings patterned via optical-projection lithography will have the front-side L2 mesh in the same mask as the CAT and L1 support, enabling a much simpler process than the process demonstrated via interference lithography and multiple mask layers.    

\begin{figure} [ht]
   \begin{center}
   \begin{tabular}{ c c } 
   \includegraphics[height=5.5cm]{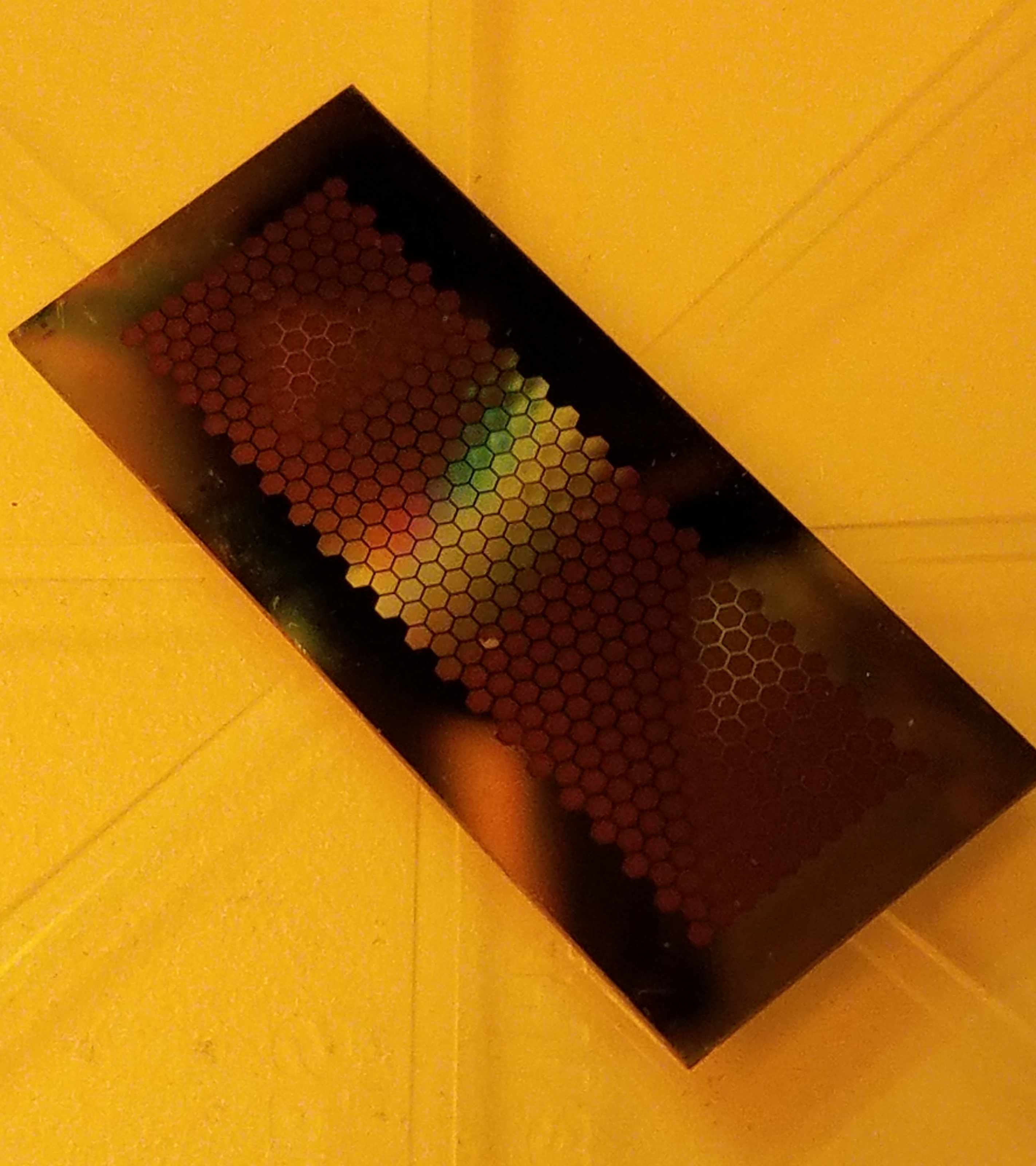}
   \includegraphics[height=5.5cm]{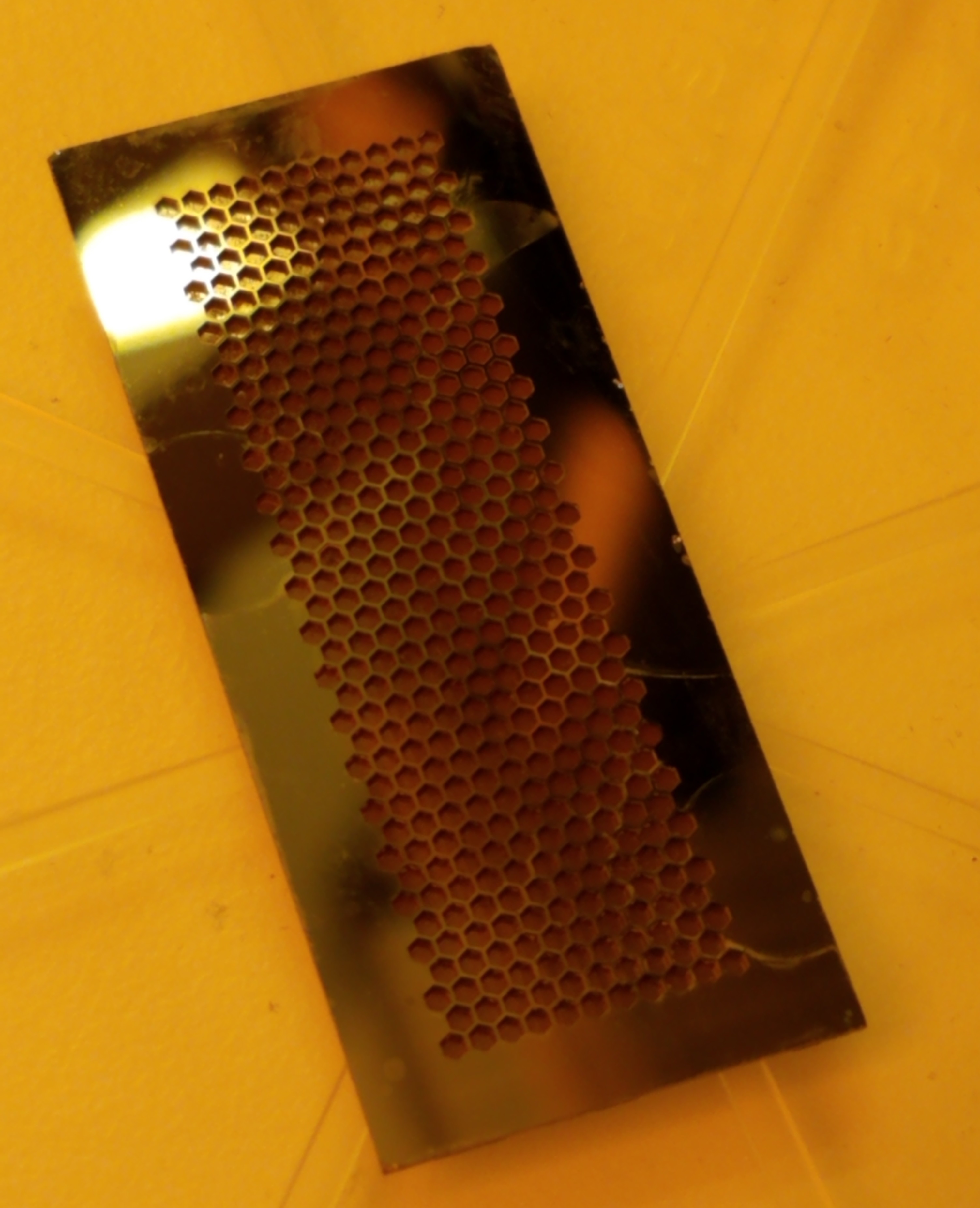}
   \end{tabular}
   \end{center}
   \caption[example] 
   { \label{fig:frontL2} 
CAT grating membrane ($\sim 10 \times 30$ mm$^2$) with aligned front (left image) and back side (right image) L2 mesh. }
\vspace{4mm}
\end{figure} 

\section{Process transfer to 200 mm state-of-the-art tools}

\begin{figure} [ht]
   \begin{center}
   \begin{tabular}{ c c } 
   \includegraphics[height=5.5cm]{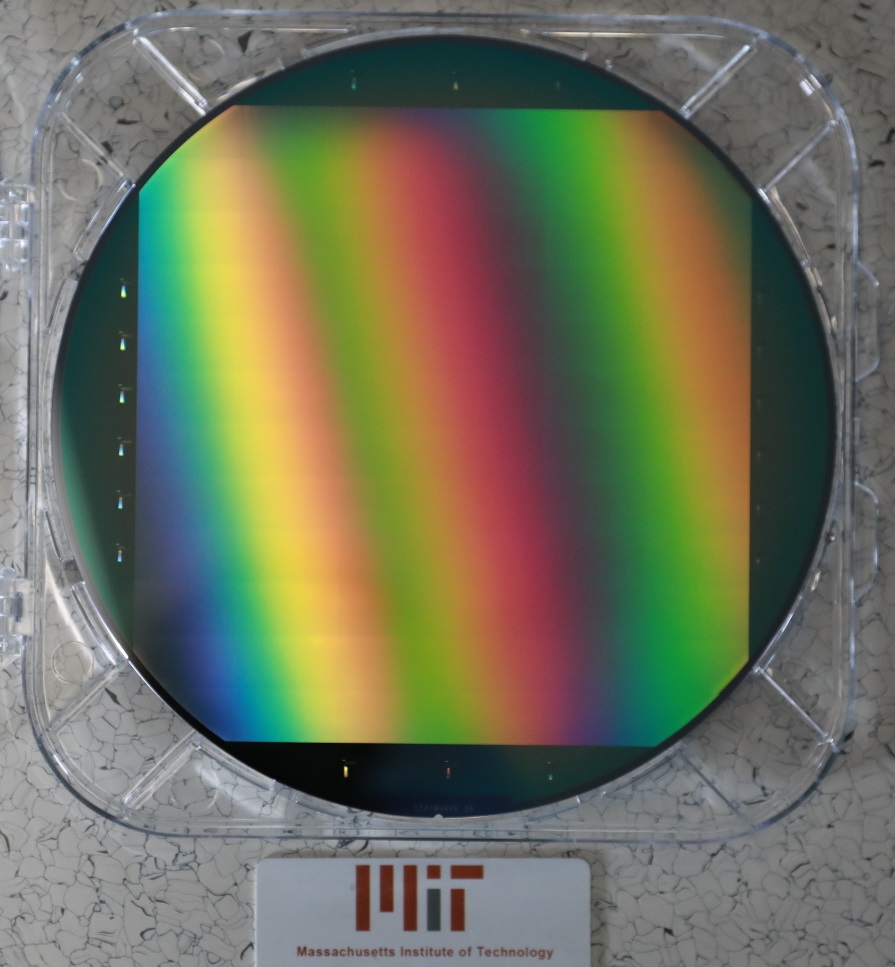}
   \includegraphics[height=5.5cm]{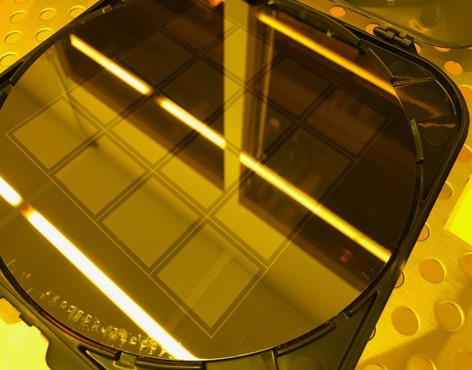}
   \end{tabular}
   \end{center}
   \caption[example] 
   { \label{fig:200mm} 
Left: 200 mm wafer front side oxide mask patterned at LL with CAT grating and L1 mesh.  The rainbow pattern is due to diffraction from the L1 mesh.  Wagon wheel patterns are visible around the perimeter.  Right: 200 mm wafer with back side L2 and L3 pattern in resist written by an MLA tool in MIT.nano.}
\end{figure}

For a mission such as Arcus or Lynx, on the order of $\sim 700-2000$ grating facets have to be manufactured with reliable and repeatable performance over a limited period of time.  In contrast, most of the gratings we have fabricated so far were made in a manual and time-consuming ``beaker and tweezer" fashion, often using shared facilities and tools with limited repeatability.  Scaling up CAT grating fabrication requires a more efficient approach with more repeatable performance, more predictable outcomes, and lower costs.  For this reason we have begun a collaboration with the MIT Lincoln Labs (LL) 
Advanced Imager Technology Group and the LL Microelectronics Laboratory.  This group has automated coating, patterning and etching tools for the batch processing of 200 mm wafers.  At the same time the new MIT.nano facility on the MIT campus is bringing new 200 mm tools online that can be used for post-LL process steps.

A key to achieving repeatable fabrication outcomes for the complex CAT grating fabrication sequence is the initial patterning step.  The grating period has to be repeatable over all lithographic exposures and constant over all areas as a necessary requirement for achieving high resolving power.  The grating etch masks must have a well-defined duty cycle to enable reliable outcomes for the challenging device layer DRIE step.  The LL ME group has a 193 nm 4X projection lithography stepper/scanner tool for 200 mm wafers.  We performed initial tests on this tool using an electron-beam (e-beam) written mask with integrated CAT grating and L1 patterns.  The tool exposure field can be stepped over the whole wafer, placing a nearly continuous and constant pattern almost everywhere. Instead of one Arcus-size grating from the center of a 100 mm wafer using interference lithography we hope to obtain 16-20 gratings out of a single 200 mm wafer (see Fig.~\ref{fig:200mm}).

LL has developed a layer stack of amorphous Si and silicon nitride to transfer the photomask pattern from the photoresist into the 300 nm thick thermal oxide layer. X-ray diffraction from the device layer has been used to measure the angle between the \{111\} planes and the wafer notch and to clock the wafer accordingly for the CAT grating line pattern to be parallel to the \{111\} planes.  The photomask also contains spoked wagon wheel patterns\cite{alex2} that have been used to verify proper clocking. 

The first few wafers patterned with this mask have mainly been used for front side DRIE development, using both four and six $\mu$m thick device layers. The mask has successfully been used to etch CAT grating bars to a depth of six $\mu$m over full 200 mm SOI wafers in a state-of-the-art DRIE tool located abroad. Preliminary work on KOH polishing has been started as well as efforts to measure the grating bar tilt angles\cite{JungkiEIPBN} along full 200 mm wafers. 

\begin{figure} [ht]
   \begin{center}
   \begin{tabular}{ c c } 
   \includegraphics[height=4.65cm]{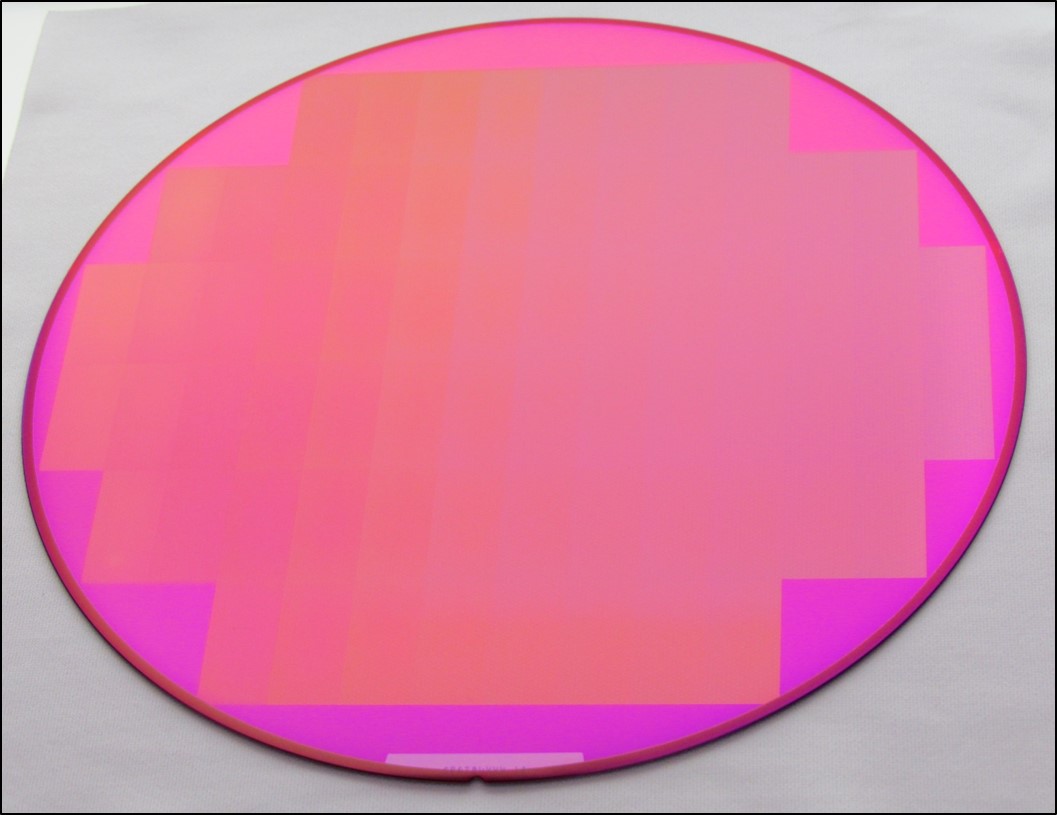}
   \includegraphics[height=4.65cm]{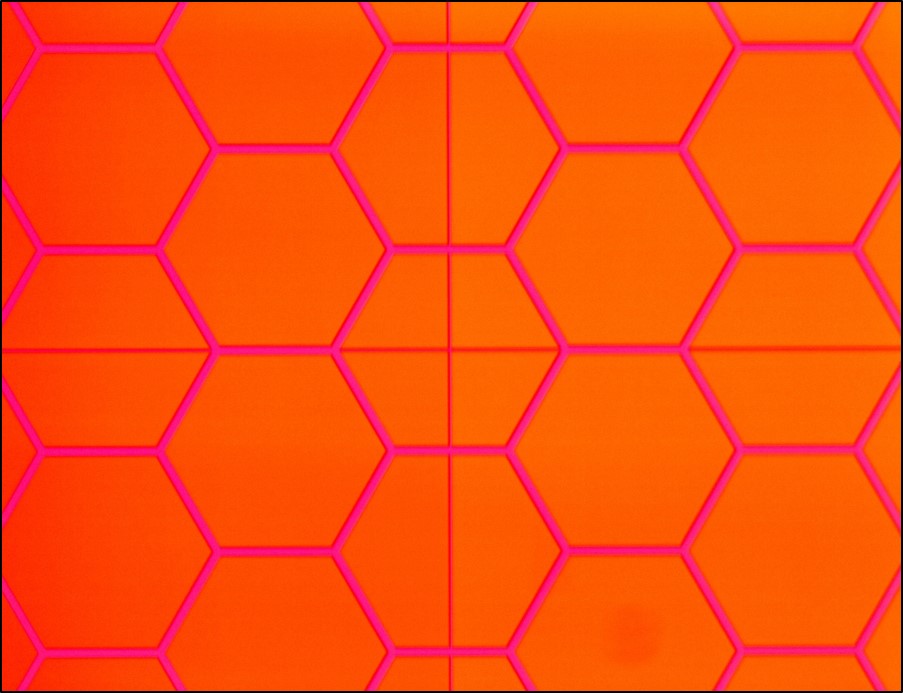}
   \includegraphics[height=4.65cm]{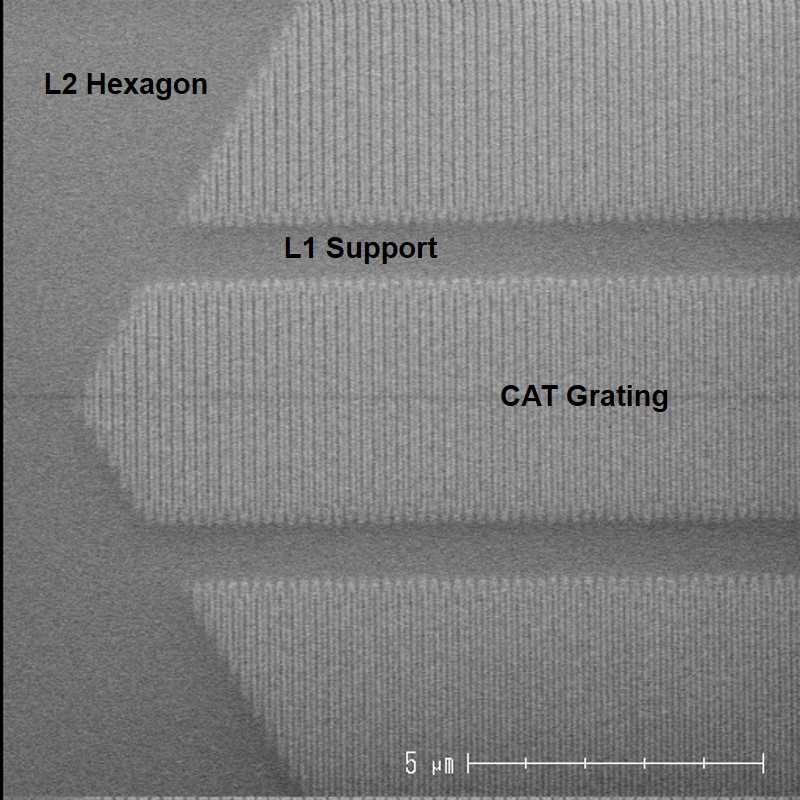}
   \end{tabular}
   \end{center}
   \caption[example] 
   { \label{fig:dose} 
Left: Result of photoresist exposure dose study across a 200 mm wafer with resist/ARC/Si$_3$N$_4$/a-Si/thermal oxide stack, using the CAT/L1/L2 front side mask.  Middle: Microscope image of the patterned resist at the intersection of four exposure fields, showing phase continuity of the ($\sim 1$ mm-pitch) L2 mesh and the solid boundaries along the field edges.  Right: Top down scanning electron micrograph, showing CAT grating and L1 and L2 support structure patterns in resist near an inside corner of the L2 hexagon mesh. (The thin dark horizontal line is an image artifact.)}
\end{figure}

Our latest photomask contains an L2 mesh (with 40 $\mu$m wide hexagon beams) in addition to the CAT grating and L1 mesh patterns (see Fig.~\ref{fig:dose}).  We have demonstrated front side patterning across a whole 200 mm wafer with phase continuity of all three patterns in both directions.  Each exposure field has a solid 5 $\mu$m boundary along its edges.  The resulting crosses at the corners of four adjacent fields can be used as alignment marks for the back side patterning of the L2 mesh (with beams wider than 40 $\mu$m). The L2 and L3 patterns are to be written by laser on the back side using a maskless aligner (MLA) tool in the MIT.nano facility. The facility is new and many of the tools are 200 mm capable. The PECVD oxide labeled in Fig.~\ref{fig:Fab} has also been etched on full 200 mm wafers via a SAMCO inductively coupled plasma etch tool. 

\section{CAT grating resolving power}

An x-ray grating spectrometer typically consists of focusing optics, an objective grating array just downstream of the optics, and a readout camera that records the dispersed source spectrum.  Assuming perfect gratings, the resolving power is to first order simply given by the diffraction angle of the order of interest, divided by the FWHM of the angular PSF of the focusing optics in the dispersion direction.  However, grating imperfections can lead to additional broadening of the spectral features, thereby reducing the resolving power.  Often this broadening is modeled as a variation of the grating period $\Delta p$.  Previously we demonstrated $R > 10{,}000$ for a $\sim 10 \times 30$ mm$^2$ CAT grating that had been patterned by interference lithography and illuminated by a single slumped glass mirror pair over an area of $\sim 1 \times 30$ mm$^2$.\cite{SPIE16,AO2019}

Interference lithography provides very smooth and well-understood phase progression over reasonably large areas, but can suffer from hyperbolic distortions, depending on the details of the setup and the desired grating size.  In the semiconductor industry patterning is primarily done using elaborate stepper/scanner tools with nm pattern placement precision over a roughly $26 \times 32$ mm$^2$ exposure field, for example.  The pattern itself is defined in a mask that is written at four times the scale of the desired patterns, using electron beam (or e-beam) lithography.  Ideally, for a grating patterned in this fashion, local and global deviations $\Delta p$ from the desired grating period $p$ should be smaller than $p/R$ over the whole exposure field.  In practice it is sufficient for the standard deviation $\sigma_p$ of the period distribution to satisfy $\sigma_p < p/R$, which is 20 pm in our case.

\begin{figure} [ht]
   \begin{center}
   \begin{tabular}{ c c } 
   \includegraphics[height=5.5cm]{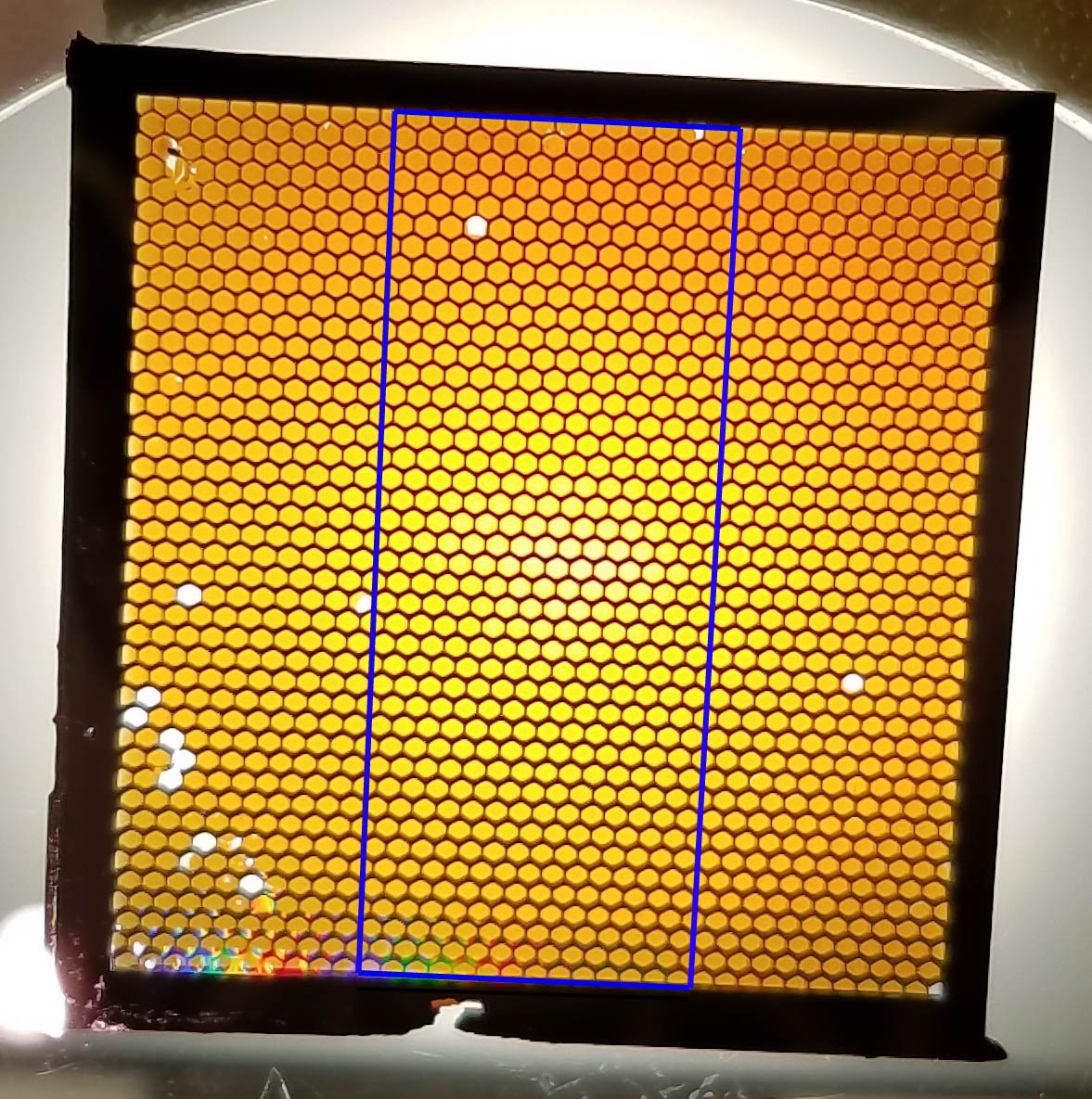}
   \includegraphics[height=5.5cm]{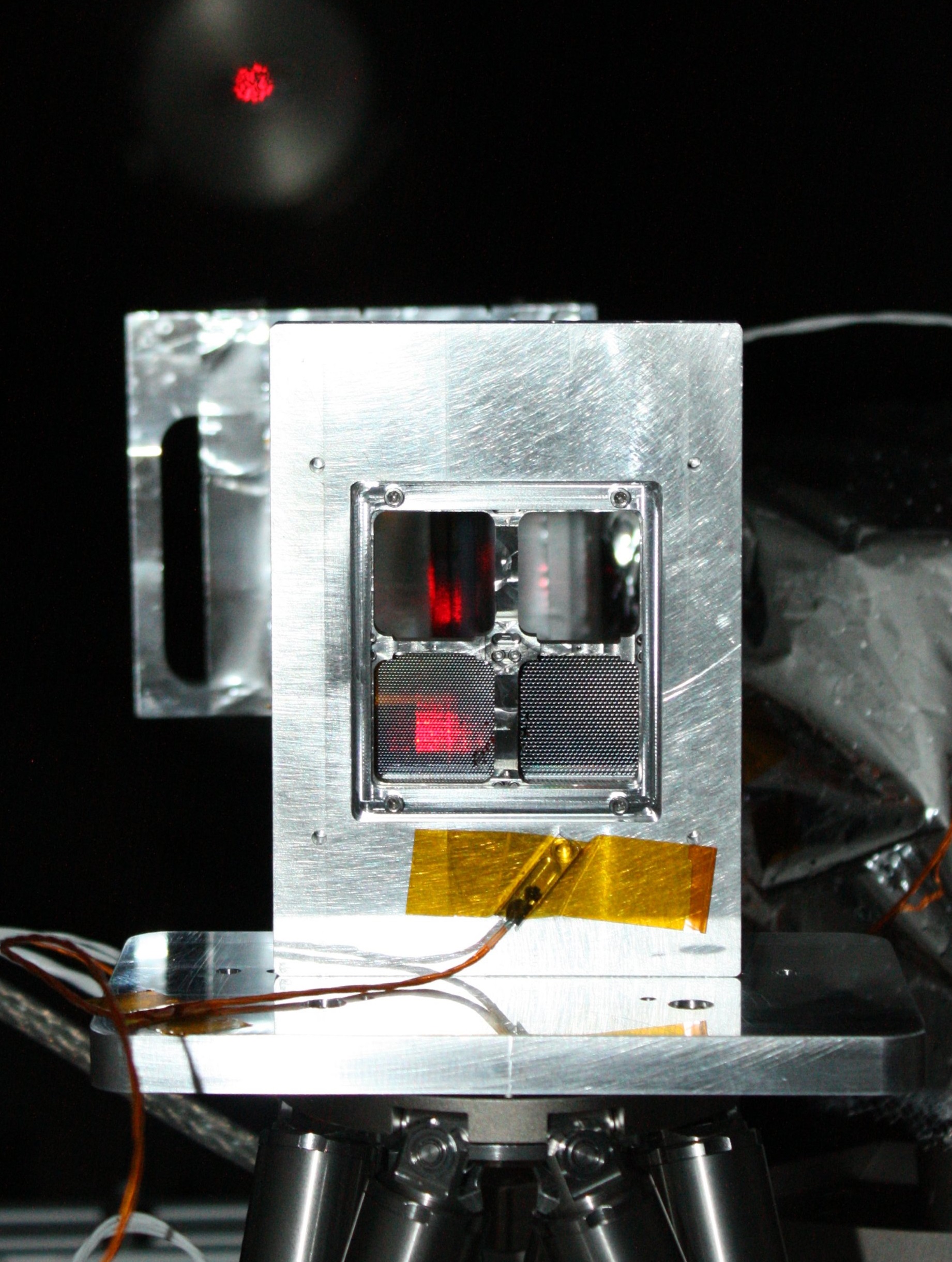}
   \includegraphics[height=5.5cm]{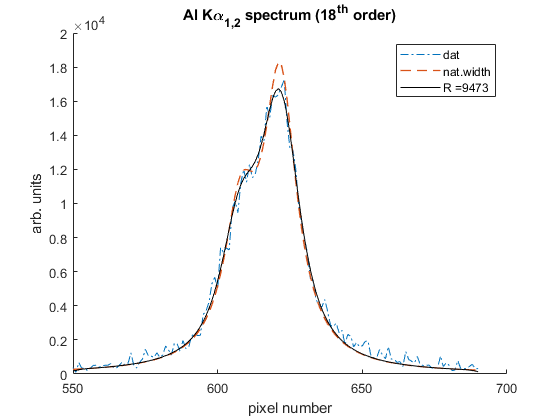}
   \end{tabular}
   \end{center}
   \caption[example] 
   { \label{fig:18th} 
Left: CAT grating membrane fabricated from a 200 mm SOI wafer with 3 $\mu$m device layer.  The outer dimensions are $\sim 30.6 \times 29.5$ mm$^2$.  The blue rectangle shows the approximate area illuminated by x rays.  Holes are mainly due to damage from manual handling.  Middle: View towards the source at PANTER during laser alignment.  The grating under test is in the lower left quadrant of the grating plate.  The upstream SPO XOU is visible through the open upper left quadrant.  The red dot in the background is an alignment laser emanating from the source end of the facility and illuminating the XOU.  Red laser light transmitted through the XOU can be seen hitting the grating.  Right: Measured K$\alpha$ spectrum in 18$^{\mathrm {th}}$ order projected onto the dispersion axis versus detector pixel number.  The data is shown with a dash-dotted line, the dashed line shows the theoretical spectrum using natural line widths, and the solid line is the fit to the data, clearly showing the shoulder in the doublet. }
\end{figure}

The most convincing way to test the grating performance is to measure the broadening due to $\sigma_p$ in an x-ray grating spectrometer.  We therefore performed a measurement very similar to our previous work,\cite{AO2019} with some important differences. 
We did not have any 200 mm SOI wafers with a $<$110$>$ device layer available at the time.  Instead we used one of two $<$100$>$ 200 mm SOI wafers with a nominally 3 micron thick device layer.  We patterned it using the LL 4X optical projection lithography stepper/scanner using an e-beam written mask with a 50\% duty cycle CAT grating pattern and an orthogonal 15\% duty cycle L1 support mesh pattern.  After projection this results in a 200 nm-period grating with a 5 $\mu$m-period orthogonal support mesh in resist.  This pattern was transferred into an underlying oxide layer, which then served as the hard mask for the DRIE through the device layer, stopping on the BOX.  Wet etching in KOH solution was not performed, since the device layer had the wrong crystal orientation.  A hexagonal L2 support mesh and an L3 frame were etched into the 625 $\mu$m thick handle layer, resulting in a silicon grating membrane with $\sim 26 \times 27$ mm$^2$ of area inside the frame.  The BOX was removed from the open areas using RIE and a vapor HF etch.  The resulting silicon grating membrane is shown on the left of Fig.~\ref{fig:18th}. The grating was then epoxied to a Ti frame at a single point in the center of one of the L3 edges.

The x-ray test was performed at the PANTER facility in Neuried, Germany (see middle of Fig.~\ref{fig:18th}).  As a focusing optic we used a 12 m-focal length SPO x-ray optical unit (XOU) with 14 stack pairs, which provided a roughly $11 \times 26.2$ mm$^2$ illumination area for the grating, and a best focus of 1.04 arcsec (FWHM) along the dispersion direction.  The grating was placed $\sim 268$ mm from the XOU node, which in turn was about 119.5 m from the electron-bombardment x-ray source.  We used an Al target, which emitted the characteristic K$\alpha_{1,2}$ doublet at 1.49 keV.  As in our previous studies, we focused our effort on the 18$^{\mathrm {th}}$
diffraction order at 4.3 degrees from the straight-through 0$^{\mathrm {th}}$ order.  The grating was rotated by half this angle for optimal blazing.  Nevertheless, the diffraction efficiency in such a high order was rather low due to two main reasons: This angle exceeds the critical angle for Si at this wavelength by a factor of $\sim 2$, and the grating bar sidewalls were rough from the DRIE.  

The data shown on the right in Fig.~\ref{fig:18th} was collected using the PIXI detector (PI-MTE 1300B, $1340 \times 1300$ 20 $\mu$m square pixels, placed 13345 mm from the mirror node due to the finite source distance) with a total exposure time of just over seven hours, with data taken on two different days simply summed up.  The data was projected onto the dispersion axis and fitted to a model of the K$\alpha_{1,2}$ doublet with parameters from Ref.~\citeonline{AO2019} that is convolved with a Gaussian of variable width, representing broadening due to all sources (source size, optics PSF, grating period variations, aberrations, drifts in the setup, etc.).  The only other fit parameters are the position of the K$\alpha_{1}$ peak and an overall scale factor.  After fitting we subtract out (in rss fashion) the measured width of the direct beam, assuming Gaussian shapes for the beam and all other sources of broadening.  We find the best-fit standard deviation of the remaining Gaussian to be $\sim 52$ $\mu$m on the detector.  If we interpret this broadening as simply due to grating period variations we find that this grating is compatible with a resolving power of $R \geq 9470$.  When we analyze the data collected on two different days separately we find $R \geq 9800$ and $R \geq 12000$, with a $\sim 10$ $\mu$m shift in best fit peak position between the two days.  At this point in our analysis it is unclear how much of this shift is statistical variation and to what degree it may present small dimensional changes in the setup over tens of hours in time. 

This result clearly shows that gratings patterned with an e-beam written mask and 4X projection lithography are suitable for high-resolution spectrographs requiring $R > 7500$.  The grating was illuminated with 14 confocal mirror pairs, and the sampled area was $\sim 10$ times larger than in our previous work.  We don't see any fundamental problems that would prevent this technology from being scaled to the size required for Arcus or Lynx.


\section{CAT grating diffraction efficiency: data and modeling}

In order to design CAT gratings for different applications it is imperative to have a good model that can predict diffraction efficiency as a function of structural and geometric parameters such as grating and coating materials, grating period, bar width and depth, and x-ray wavelength and angle of incidence.  We have used rigorous coupled-wave analysis (RCWA) for most of our modeling.  The grating structure is modeled as a sequence of stratified layers, with distinct regions of different index of refraction within each grating layer.  Periodic boundary conditions are assumed in the direction of the grating vector.  Thus grating bars are modeled with a rectangular profile (for example, width of 60 nm, depth of 4 $\mu$m) in the case of a single grating layer.  Other bar profiles can be approximated as ``pyramids" of rectangles comprised of many layers.\cite{OE,AO}

\begin{figure} [ht]
   \begin{center}
   \begin{tabular}{ c c } 
   \includegraphics[height=6cm]{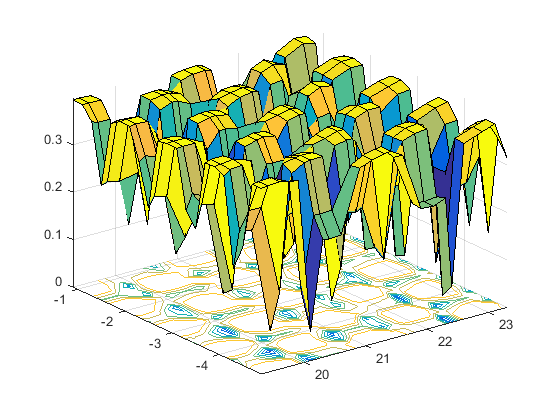}
   \includegraphics[height=6cm]{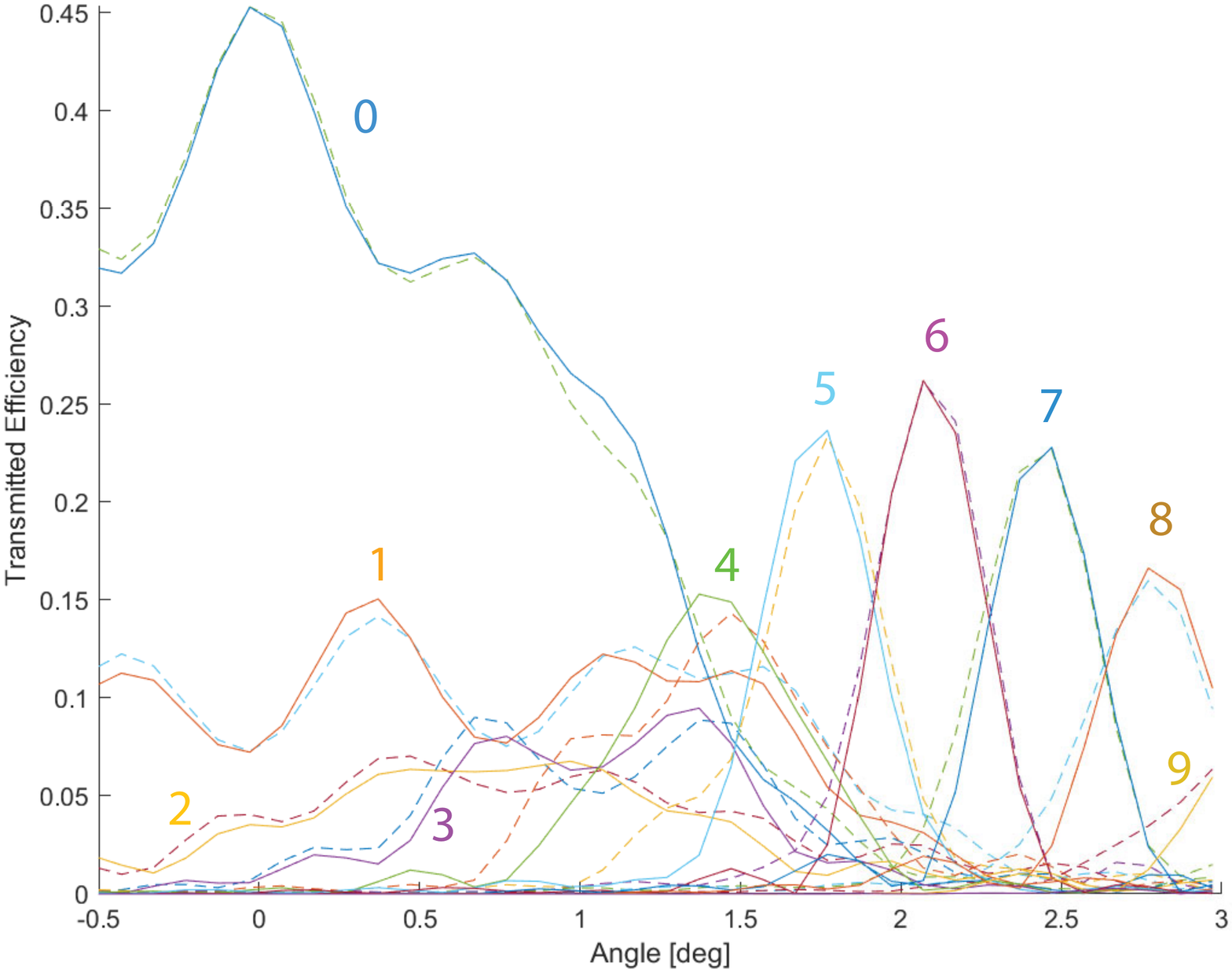}
   \end{tabular}
   \end{center}
   \caption[example] 
   { \label{fig:X14_data} 
Synchrotron data from grating X14 at an x-ray wavelength of 2.38 nm.  Left: Sum of absolute efficiencies of orders 3-8    as a function of beam footprint position (in mm) on the grating at an incidence angle of 1.7 degrees, including L1 and L2 blockage.  Clearly, the impact of the hexagonal L2 support mesh can be seen on the mm-scale.  Right: Absolute diffraction efficiency (including L1 blockage) for orders zero through nine on a single location on the grating as a function of incidence angle (``rocking scan").  Dashed lines are measured efficiencies and solid lines are model efficiencies.  See text for details.}
\end{figure}

Most of our diffraction efficiency measurements to date have been performed at the calibration and standards beamline 6.3.2 at the Advanced Light Source at Lawrence Berkeley National Laboratory.  Many measurements have been published previously, typically in the range of 80-100\% of predictions for perfect structures, but often lower closer to the critical angle or critical wavelength.\cite{SPIE16,SPIE08,SPIE09,SPIE14,SPIE15,SPIE17}  We typically sample gratings with a $\sim 50 \times 250$ $\mu$m$^2$ beam size in areas between the L2 supports.  A footprint of this size averages over many tens of L1 support bars and therefore properly reflects their impact on grating effective area.  For mission purposes and effective area predictions we also need to take blockage by L2 structures into account.  This could simply be calculated from the lithographically defined dimensions of the L2 mask.  However, pattern transfer and deep etch imperfections could lead to deviations from the theoretically expected structure.  In a recent experiment we therefore stepped gratings across the x-ray beam in many small (50-200 $\mu$m) steps over areas containing $\sim 20$ L2 hexagons and measured the diffraction efficiency for several orders at each spot at a few angles of incidence.  The left of Fig.~\ref{fig:X14_data} shows the sum of absolute efficiencies of orders 3-8 as a function of position on the grating for the angle of incidence with the highest efficiency.  Clearly the blockage from the hexagonal L2 mesh can be seen as a hexagonal pattern of reduced efficiency.  Nevertheless, the average absolute efficiency over this $\sim 4\times 4$ mm$^2$ area is 31.6\% (including L1 and L2 blockage), which already exceeds the Arcus goal of 28\%.  If we divide the average efficiency over some area by the highest efficiency found over the same area we obtain values in the range of 74-79\%, which is consistent with the L2 design open area of 81\%.

The grating bar profile cannot be discerned in sufficient detail from SEM images alone to predict diffraction efficiency with better than $\sim 10$\% relative accuracy.  We are investigating how well we can model fabricated CAT grating bar profiles from rocking scans of zeroth order at a single x-ray wavelength of interest and predict diffraction efficiency for higher orders.  The right of Fig.~\ref{fig:X14_data} shows example data (dashed lines) from grating X14 where we use a three-layer fit to the zeroth order rocking scan to find grating bar values for each layer (layer depth and grating bar width).  The effect of L1 supports is modeled with a single number, representing the percentage of grating area occupied by the L1 mesh (L1 duty cycle).  The L2 mesh is not considered, since this data was taken near the center of an L2 hexagon.  The result of the zeroth order fit is then used to model the efficiency of the higher diffraction orders.  RCWA assumes perfectly rectangular profiles for each grating layer and predicts higher efficiencies than the measured ones.  We model grating bar imperfections such as roughness heuristically as a single wavelength- and angle-dependent Debye-Waller factor $f_{\mathrm {DW}}$, which has no impact on zeroth order, but decreases the efficiency of higher orders according to

\begin{equation}
    f_{\mathrm {DW}} = \exp{(-q_z^2 \sigma^2)} = \exp{ \left[- \left({2\pi m \sigma \over{p}}\right) ^2 \right]},
    \label{DWG}
\end{equation}

\noindent
where $q_z = q_z^{out} - q_z^{in} = (2\pi/\lambda)(\sin\beta_m - \sin\theta)$ is the wavevector transfer parallel to the grating vector and $\sigma$ is a roughness parameter.  The solid lines on the right of Fig.~\ref{fig:X14_data} were generated using $\sigma = 2.6$ nm, which is a reasonable number if we chose to interpret it as rms sidewall roughness. The total depth of this grating is found to be 4.0 $\mu$m, and the grating bar widths range from 43 nm at the side facing the x-ray source to 38 nm on the side facing the detector.  The L1 duty cycle is found to be 16\%.  The agreement between data and model is quite good over many orders and a wide range of incidence angles, giving us confidence in our modeling approach.

\section{Discussion and outlook}

CAT grating technology has progressed significantly over the last decade.  We believe its TRL is close to six for the purposes of the Arcus mission.  Arrays of up to four aligned gratings have been built and shown to perform within expectations.\cite{SPIE17,SPIE18}  Our fabrication approach is relatively mature, but we continue to look for improvements that lead to shorter fabrication times and better and more reliable outcomes.  As an example,
the addition of a front side L2 mesh, aligned with the back side L2 mesh, leads to wider process latitude and higher mechanical strength.  Once added to the photomask pattern it requires no additional fabrication steps (besides alignment of the back side L2 mesh).

Manufacturing hundreds or thousands of CAT gratings demands a more advanced tool set, larger wafers, and a larger degree of automation and repeatability.  Transferring a complex multi-step nanofabrication process to a different tool set is not trivial.  The demands for process uniformity across 200 mm wafers are challenging, even for state-of-the-art tools.  Nevertheless, we have made steady progress in moving from 100 mm wafers processed on research tools open to the whole campus community to processing 200 mm wafers on more restricted and better controlled production style tools.  Optical 4X projection lithography with commercially e-beam written masks has produced grating areas of significant size with resolving power on the order of $10^4$.  To the best of our knowledge such x-ray grating resolving power has never been shown before in combination with e-beam patterning.

The use of a mask with a fixed, precise angle between CAT grating bars and L1 mesh bars opens up the opportunity for high-precision roll alignment between gratings using visible-light diffraction from the L1 mesh.  This can be done in a normal incidence transmission geometry, which is much less sensitive to grating non-flatness than methods based on reflection.\cite{JungkiSPIE2018}  We are currently exploring this alignment method.

The back side pattern defines the back side L2 mesh, which is aligned to the front side L2 mesh, which in turn has a fixed orientation with respect to the CAT grating bars.  It also defines the L3 frame of a grating membrane with lithographic precision.  This means we can fabricate macroscopic precision structures (the outside dimensions of the L3 frame) with a precisely defined orientation relative to the nm-scale CAT grating bars.  In principle our fabrication scheme makes it possible to use the L3 edges as reference surfaces for fast and precise grating alignment and integration into large arrays.

We expect e-beam patterning of masks for chirped gratings to be feasible as well.  The ability to chirp would allow us to make larger (and thus fewer) gratings for Lynx, without introducing unacceptable losses in $R$ caused by aberrations due to the deviation of flat gratings from the Rowland torus surface.\cite{CAT-JATIS,moritz2020}

We continue to work on transferring more of the fabrication steps onto 200 mm tools for faster and cheaper production.  Process uniformity over the large area of a 200 mm wafer remains a challenge and may limit the number of quality gratings that can be extracted from a single wafer.

Individual CAT grating performance exceeds currently envisioned resolving power requirements for both Arcus and Lynx.  Diffraction efficiency meets Arcus requirements.  Lynx demands higher diffraction efficiency, which can be achieved by moving from 4 $\mu$m to 6 $\mu$m grating depth.  We are well on our way toward this goal.  CAT grating DRIE down to 6 $\mu$m depth on SOIs has been accomplished, and we have begun KOH polishing on those gratings.  In addition, for Lynx the fraction of grating area blocked by L1 and L2 structures also needs to be reduced from 15-20\% to the 10\% level per mesh while still providing enough mechanical strength.  As part of this work we have performed detailed investigations of alternative L2 structures using analytical and finite-element approaches.\cite{Jungkithesis}

We are progressing toward a situation where we can produce a significant number of CAT gratings with very similar structural parameters, which will allow us to perform more systematic experiments and process development.  Hopefully this will also provide us with a sufficient number of gratings to test the performance of progressively larger grating arrays that can serve as believable surrogates for fully sized proposed instruments.

\acknowledgments 
 
We gratefully acknowledge facility support from Microsystems Technology Labs, the Nanostructures Lab and MIT.nano, all at MIT, and SAMCO, Inc.  We thank Eric Gullikson for support at the ALS.  A part of this work used resources of the Advanced Light Source, which is a DOE Office of Science User Facility under contract no. DE-AC02-05CH11231.  Part of the work performed at PANTER has been supported by the European Unions Horizon 2020 Programme under the AHEAD project (grant agreement n.~654215).
This work was supported by NASA grants NNX17AG43G, 80NSSC19K0335, and 80NSSC20K0780.


\begin{thebibliography}{1}

\bibitem{Arcus2019}  Smith, R. K., {\it et al.}
``Arcus: The soft x-ray grating Explorer,"
{\it Proc. SPIE} {\bf 11118}, 11118W0 (2019).

\bibitem{Lynx2019}  Bautz, M. W., 
``The Lynx x-ray observatory: Science drivers,"  
{\it Proc. SPIE} {\bf 11118}, 111180J (2019).

\bibitem{cxc}  C.~R.~Canizares {\it et al.}, ``The Chandra high-energy transmission grating: Design, fabrication, ground calibration, and 5 years in flight," 
PASP {\bf 117}, 1144-1171 (2005).

\bibitem{RGS} J.~W.~den Herder {\it et al.}, ``The reflection grating spectrometer on board XMM-Newton,"
Astr. \& Astroph. {\bf 365}, L7-L17 (2001).

\bibitem{Collon2018} Collon, M. J., {\it et al.},
``Silicon pore optics mirror module production and testing,"
{\it Proc. SPIE} {\bf 11180}, 1118026 (2019).

\bibitem{Athena} \url{http://sci.esa.int/athena/}

\bibitem{SPIE18} Heilmann, R.~K., {\it et al.},
``Blazed transmission grating technology development for the Arcus x-ray spectrometer Explorer ,"
{\it Proc. SPIE} {\bf 10699}, 106996D (2018).

\bibitem{moritz2017} G\"unther, H.~M., Heilmann, R. K., Cheimets, P., and Smith, R. K.,
``Performance of a double tilted-Rowland-spectrometer on Arcus,"
{\it Proc. SPIE} {\bf 10397}, 103970P (2017).

\bibitem{moritz2018} G\"unther, H.~M.~{\it et al.},
``Ray-tracing Arcus in Phase A,"
{\it Proc. SPIE} {\bf 10699}, 106996F (2018).

\bibitem{XRISM} Tashiro, M. {\it et al.} 
``Concept of the X-ray Astronomy Recovery Mission," 
{\it Proc. SPIE} {\bf 10699}, 1069922 (2018).

\bibitem{Gaskin2019} Gaskin, J. A., {\it et al.},
``Lynx x-ray observatory: An overview,"
{\it J. Astron. Telesc. Instrum. Syst.} {\bf 5}, 021001 (2019).

\bibitem{CAT-JATIS} G\"unther, H.~M., and Heilmann, R. K.,
``Lynx soft x-ray critical-angle transmission grating spectrometer,"
{\it J. Astron. Telesc. Instrum. Syst.} {\bf 5}, 021003 (2019).

\bibitem{moritz2019} G\"unther, H.~M., and Heilmann, R. K.,
``Design progress on the Lynx soft x-ray critical-angle transmission grating spectrometer,"
{\it Proc. SPIE} {\bf 11118}, 111181C (2019).

\bibitem{Lynxreport} \url{https://www.lynxobservatory.com/}

\bibitem{CATroadmap} \url{https://wwwastro.msfc.nasa.gov/lynx/docs/documents/TechnologyRoadmaps/CAT_TR.pdf}

\bibitem{SPIE19} Heilmann, R.~K., Bruccoleri, A. R., Song, J., and Schattenburg, M. L.,
``Progress in x-ray critical-angle transmission grating technology development,"
{\it Proc. SPIE} {\bf 11119}, 1111913 (2019).

\bibitem{metashells} Zhang, W.~W., {\it et al.},
``High-resolution, lightweight, and low-cost x-ray optics for the Lynx observatory,"
{\it J. Astron. Telesc. Instrum. Syst.} {\bf 5}, 021012 (2019).

\bibitem{alex2} Bruccoleri, A.~R., Guan, D., Mukherjee, P., Heilmann, R.~K., Schattenburg, M.~L. and Vargo, S.,
``Potassium hydroxide polishing of nanoscale deep reactive-ion etched ultra-high aspect ratio gratings,"
{\it J. Vac. Sci. Technol. B} {\bf 31}, 06FF02 (2013).

\bibitem{EIPBN2016} Bruccoleri, A.~R., Heilmann, R.~K., and Schattenburg, M.~L.,
``Fabrication process for 200 nm-pitch polished freestanding
ultra-high aspect ratio gratings,"
{\it J. Vac. Sci. Technol. B} {\bf 34}, 06KD02 (2016).

\bibitem{JungkiEIPBN} Song, J., Heilmann, R.~K. and Schattenburg, M.~L., ``Characterizing profile tilt of nanoscale deep-etched gratings via x-ray diffraction,"
{\it J. Vac. Sci. Technol. B} {\bf 37}, 062917 (2019).

\bibitem{SPIE16} Heilmann, R.~K., Bruccoleri, A.~R., Kolodziejczak, J., Gaskin, J.~A., O’Dell, S.~L., Bhatia, R., and Schattenburg, M.~L.,
``Critical-angle x-ray transmission grating spectrometer with extended bandpass and resolving power $> 10{,}000$ ,"
{\it Proc. SPIE} {\bf 9905}, 99051X (2016).

\bibitem{AO2019} Heilmann, R.~K., Kolodziejczak, J., Bruccoleri, A. R.,  Gaskin, J. A., and Schattenburg, M.~L.,
``Demonstration of resolving power $\lambda /\Delta\lambda > 10{,}000$ for a space-based x-ray transmission grating
spectrometer,"
{\it Appl. Opt.} {\bf 58}, 1223-1238 (2019).

\bibitem{OE} Heilmann, R.~K., Ahn, M., Gullikson, E.~M. and Schattenburg, M.~L., ``Blazed high-efficiency x-ray diffraction via transmission through arrays of nanometer-scale mirrors,"
{\it Opt. Express} {\bf 16}, 8658-8669 (2008).

\bibitem{AO} Heilmann, R.~K., Ahn, M., Bruccoleri, A., Chang, C.-H., Gullikson, E.~M., Mukherjee, P. and Schattenburg, M.~L.,
``Diffraction efficiency of 200 nm period critical-angle transmission gratings in the soft x-ray and extreme ultraviolet wavelength bands,"
{\it Appl. Opt.} {\bf 50}, 1364-1373 (2011).

\bibitem{SPIE08} Heilmann, R.~K., Ahn, M. and Schattenburg, M.~L., ``Fabrication and performance of blazed transmission gratings for x-ray astronomy,"
{\it Proc. SPIE} {\bf 7011}, 701106 (2008).

\bibitem{SPIE09} Heilmann, R.~K., {\it et al.}, ``Development of a critical-angle transmission grating spectrometer for the International X-Ray Observatory,"
{\it Proc. SPIE} {\bf 7437}, 74370G (2009).

\bibitem{SPIE14} Heilmann, R.~K., Bruccoleri, A.~R., Guan, D. and Schattenburg, M.~L., ``Fabrication of large-area and low mass critical-angle x-ray transmission gratings,"
{\it Proc. SPIE} {\bf 9144}, 91441A (2014).

\bibitem{SPIE15} Heilmann, R.~K., Bruccoleri, A.~R. and Schattenburg, M.~L., ``High-efficiency blazed transmission gratings for high-resolution soft x-ray spectroscopy,"
{\it Proc. SPIE} {\bf 9603}, 960314 (2015).

\bibitem{SPIE17} Heilmann, R.~K., {\it et al.},
``Critical-angle transmission grating technology development for high resolving power soft x-ray spectrometers on Arcus and Lynx,"
{\it Proc. SPIE} {\bf 10399}, 1039914 (2017).

\bibitem{JungkiSPIE2018} Song, J., Heilmann, R.~K., Bruccoleri, A.~R. and Schattenburg, M.~L.,
``Metrology for quality control and alignment of CAT grating spectrometers,"
{\it Proc. SPIE} {\bf 10699}, 106990S (2018).

\bibitem{moritz2020} G\"unther, H.~M., and Heilmann, R. K.,
``Lynx grating spectrometer design: Optimizing chirped transmission gratings,"
these proceedings.

\bibitem{Jungkithesis} Song, J.,
``Metrology and mechanics for manufacturing space-based x-ray grating spectrometers,"
Ph.~D.~thesis, Dept. of Mechanical Engineering, Massachusetts Institute of Technology (2021).




\end{thebibliography}

\end{document}